\newcommand{\ben}{\begin{enumerate}}
\newcommand{\een}{\end{enumerate}}
\newcommand{\ba}{\begin{array}}
\newcommand{\ea}{\end{array}}
\newcommand{\be}{\begin{equation}}
\newcommand{\ee}{\end{equation}}
\newcommand{\bea}{\begin{eqnarray}}
\newcommand{\eea}{\end{eqnarray}}

\documentclass[a4paper,11pt]{article}
\pdfoutput=1 % if your are submitting a pdflatex (i.e. if you have
\usepackage{jheppub} % for details on the use of the package, please
\usepackage[T1]{fontenc} % if needed

\title{\boldmath Quasi-Integrability of The KdV System}

\author[a]{K. Abhinav}
\author[a,1]{Partha Guha\note{Corresponding author.}}

\affiliation[a]{S N Bose National Centre for Basic Sciences, JD Block, Sector III, Salt Lake, Kolkata 700106,  India}

\emailAdd{kumar.abhinav@bose.ac.in}
\emailAdd{partha@bose.res.in}

\abstract{The quasi-integrable KdV equation has been obtained from the corresponding deformation of the Hamiltonian for
the usual KdV system. Following suitable gauge-fixing, it has been found that the quasi-conservation condition is satisfied and an infinite number
of anomalous conservation laws are obtained, with some containing possible conserved charges. Judicious choice of deformation 
of the Hamiltonian clearly leads to a conserved charge, manifesting quasi-integrability, but also creates a hierarchy of
higher-derivative equations with at least one conserved charge. A particular quasi-deformation parameterization of the Hamiltonian
is found to complement the same of the NLS system, following an approximate equivalence of the two systems obtained earlier, in 
the weak coupling limit. Single-soliton solutions for all these cases are obtained, with manifest scaling due to the
quasi-integrable deformation. Finally, quasi-conservation formulation for the complex coupled generalized KdV system is obtained.}

\begin{document} 
\maketitle
\flushbottom

\section{Introduction}
The Korteweg-de Vries (KdV) equation and the Nonlinear Schr\"odinger (NLS) equation are the best known
$1+1$-dimensional integrable partial differential equations (PDEs) \cite{2}. However, their mathematical natures are different. In fact, 
KdV is geometrically connected to diffeomorphism group \cite{Arnold} whereas NLS is tagged with loop algebra \cite{Chau}. 
Moreover the Lax representation \cite{Lax} of KdV system involves second and third order monic differential operators $(L,A)$,
whereas the Lax representation of NLS is given by $2\times 2$ matrices \cite{2}. However, it is known that using a particular
ansatz, in the  suitable weak-coupling limit, the nonlinear Schr\"odinger equation can be approximated by the KdV
equation \cite{5,Jun}. 

Real physical systems are characterized by finite number of degrees of freedom, prohibiting integrability of the
corresponding field-theoretical models in principle. However, they are physically known to posses solitonic states, very
similar in structure to integrable ones, like sine-Gordon (SG) \cite{FerrSG}. This motivates the study of physical continuous
systems with the interpretation as slightly deformed integrable models. In a recent work \cite{FZ,FZ1}, the SG model was
shown to be deformable into an approximate system, characterized by a {\it finite} number of conserved charges. The corresponding
connection was almost flat, yielding an anomalous zero-curvature condition. The system was thus deemed quasi-integrable
(QI). 

\smallskip

In a very interesting paper Ferreira et al \cite{FZ2}
considered modifications of the NLSE to investigate the
recently introduced concept of quasi-integrability, where they modified NLS potential
of the form $V\approx(|\psi|^2)^{2 + \epsilon}$, with $\epsilon$
being a perturbation parameter, and showed that such
models possess an infinite number of quasi-conserved charges.
Recently, a connection between non-holonomic (NH) and quasi-integrable (QI) deformations, subjected to 
the NLS system, has been obtained \cite{AGM}. This is of interest owing to the fact
that QI systems retain integrability in the asymptotic limit, whereas NH deformation maintains integrability generically
\cite{FZ}. 

\bigskip

In this paper we obtain the quasi-integrable deformation of the KdV equation. It is known that
the Lax representation of KdV involves second and third order monic differential operators $(L,A)$ and 
on the other hand Lax representation of NLS is given by $2\times 2$ matrices \cite{2}. Unfortunately the 
traditional Lax representation of the KdV equation cannot be applied to obtain the quasi-integrable deformation because of the
appearance of spurious terms. So we use loop algebra representation of the KdV equation to derive the 
quasi-integrable deformation. However, KdV being a more than second order differential system, the equation of motion
does not leave room for a {\it dynamic} deformation of the Lax pair, as done for SG \cite{FZ,FZ1} and NLS \cite{FZ2,AGM}
systems. Instead, an {\it off-shell} deformation, that of the KdV Hamiltonian has been achieved, allowing for specific
deformation of the suitable Lax component, interestingly yielding a hierarchy of higher-derivative extensions of KdV
which are QI in nature, including a scaled version of KdV. In the perturbative domain, as it is known that using a
certain ansatz the nonlinear Schr\"odinger equation can be approximated by the KdV equation \cite{5,Jun}, a particular
type of the said QI deformation of KdV was directly mapped to QI NLS systems, extending the validity of our result. Using
this connection, and from the preceding general treatment, we obtain a general structure for the quasi-integrable
deformation of KdV system and recover the results similar to that of Ferreira {\it et al.} \cite{FZ2} for NLS system;
however, through an independent approach owing to the absence of definite parity-resolution in the solution-space, unlike 
that in case of QI NLS system.

\section{Quasi-Integrable Deformation of KdV equation}

\subsection{Zakharov-Sabat representation}
From the Adler, Kostant and Symes (AKS) equation, it is possible to obtain a pair of coupled complex KdV equations
\cite{1}, through construction of the Lax pair:

\be
A=Q \quad{\rm and}\quad B=T+[S,Q],\label{1}
\ee
where,

\bea
&&T=-Q_{xx}+\left[Q^+,\left[Q^-,Q^+\right]\right]-\left[Q^-,\left[Q^-,Q^+\right]\right] \quad{\rm and}\nonumber\\
&&S=Q^+_x+Q^-_x+4c\left(Q^++Q^-\right),\qquad c\in{\mathbb R},\label{2}
\eea
with the definitions,

\bea
&&Q=\left( \ba{cc} 0 & \bar{q} \\ -q & 0 \ea \right);\quad Q^+=\left( \ba{cc} 0 & \bar{q} \\ 0 & 0 \ea \right),\quad Q^-=\left( \ba{cc} 0 & 0 \\ -q & 0 \ea \right),\label{3N}\\
&&Q=Q^++Q^-\equiv \bar{q}\sigma_+-q\sigma_-.\nonumber
\eea
where, the Pauli matrices satisfy the $SU(2)$ algebra:

\be
[\sigma_+,\sigma_-]=\sigma_3=\left( \ba{cc} 1 & 0 \\ 0 & -1 \ea \right) \quad{\rm and}\quad [\sigma_3,\sigma_\pm]=\pm 2\sigma_\pm,\label{4}
\ee
and $q,~\bar{q}$ are mutually complex-conjugate amplitudes, leading to the coupled KdV equations. It is easy to see that
an $sl(2)$-loop algebra can be constructed on the $SU(2)$ basis, leading to a complete gauge-group interpretation of this
system.

\paragraph*{}Incorporating the generic representation of Eq.s \ref{3N}, the Lax pair takes the form,

\bea
&&A=\bar{q}\sigma_+-q\sigma_- \quad{\rm and}\nonumber\\
&&B=\left(\bar{q}q_x-q\bar{q}x\right)\sigma_3-\left(\bar{q}_{xx}+2\vert q\vert^2\bar{q}\right)\sigma_++\left(q_{xx}+2\vert q\vert^2q\right)\sigma_-, \quad \vert q\vert^2=q\bar{q}.\label{23}
\eea
The corresponding curvature then can be evaluated as,

\be
F_{tx}=\left(\bar{q}_t+\bar{q}_{xxx}+6\vert q\vert^2\bar{q}_x\right)\sigma_+-\left(q_t+q_{xxx}-6\vert q\vert^2q_x\right)\sigma_-,\label{24}
\ee
yielding two coupled KdV-like equations,

\be
\bar{q}_t+\bar{q}_{xxx}+6\vert q\vert^2\bar{q}_x=0 \quad{\rm and}\quad q_t+q_{xxx}-6\vert q\vert^2q_x=0,\label{25}
\ee
under zero-curvature condition, by considering each linearly independent component of the curvature matrix. Although the
above equations posses higher order non-linearity than the usual KdV system, a straight-forward choice of variables, 

\be
\bar{q}=u, \quad{\rm and}\quad q=1;\qquad u\in{\mathbb R},\label{5}
\ee
immediately leads to the \underline{non-coupled} (usual) KdV equation,

\be
u_t+6uu_x+u_{xxx}=0.\label{8}
\ee
The other possibility: $\bar{q}=1,~q=u$ leads to a KdV equation with a negative sign to the non-linear
term, which can be transformed to the `usual' one through the transformation $u\rightarrow -u$. The Lax pair corresponding
to the choice in Eq. \ref{5} is,

\bea
&&A=u\sigma_+-\sigma_- \quad{\rm and}\nonumber\\
&&B=-u_x\sigma_3-\left(u_{xx}+2u^2\right)\sigma_++2u\sigma_-,\label{6}
\eea
leading to the curvature,

\be
F_{tx}:=A_t-B_x+[A,B]\equiv\left(u_t+6uu_x+u_{xxx}\right)\sigma_+.\label{7}
\ee

\subsection{Quasi-integrable Deformation}Since the KdV equation has derivatives higher than two, a dynamic interpretation 
of the same is not possible at the level of the equation itself. In order to employ the quasi-integrability mechanism of 
Ref.s \cite{FZ,FZ1,FZ2,N1}, the notion of {\it potential} is essential, that emerges from such an interpretation of the equation.
In case of KdV system, however, a well-known Hamiltonian formulation \cite{2} exists. In fact, the KdV equation \ref{8}
can be shown to emerge from two different equivalent Hamiltonians. Subjected to the order of non-linearity appearing in
the Lax pair of Eq. \ref{6}, we opt for the following Hamiltonian,

\be
H_1[u]=\int_{-\infty}^\infty dx~\left(\frac{1}{2}u_x^2-u^3\right) \quad{\rm with}\quad \frac{\delta H_1[u]}{\delta u(x)}=-3u^2-u_{xx}.\label{9}
\ee
This enables us to re-express the time component ($B$) of the Lax pair as,

\be
B\equiv-u_x\sigma_3-\left[u_{xx}-\frac{2}{3}\left(\frac{\delta H_1[u]}{\delta u(x)}+u_{xx}\right)\right]\sigma_++2u\sigma_-.\label{11}
\ee 
The above is a general expression to accommodate any possible deformation at the Hamiltonian level. We propose that the 
deformation of the system is implemented in the non-linear part of Hamiltonian for the KdV system to impart
quasi-integrability, the explicit form of which will be discussed below. The corresponding curvature takes the form:

\be
F_{tx}\equiv\left[u_t+u_{xxx}-\frac{2}{3}\partial_x\left(\frac{\delta H_1[u]}{\delta u(x)}+u_{xx}\right)+2uu_x\right]\sigma_++{\cal X}\sigma_3,\label{12}
\ee
with the supposed anomaly term,

\be
{\cal X}=2u^2+\frac{2}{3}\left(\frac{\delta H_1[u]}{\delta u(x)}+u_{xx}\right),\label{13}
\ee
that vanishes for undeformed system\footnote{\noindent One can very well work with the {\it second} Hamiltonian form for the KdV
system \cite{2}: $H_2[u]=-\int_{-\infty}^\infty dx~\frac{1}{2}u^2(x)$, with the alternate fundamental bracket defined as
$\left\{u(x),u(y)\right\}=\left[\partial^3_x+2\left(u_x+u\partial_x\right)\right]\delta(x-y)$. Then, the time component 
of the Lax pair will take the form: $B\equiv-u_x\sigma_3-\left[u_{xx}-4\frac{\delta H_2[u]}{\delta x}\right]\sigma_++2u\sigma_-$.
Rest will follow through the replacement: 
$\frac{2}{3}\left(\frac{\delta H_1[u]}{\delta u(x)}+u_{xx}\right)\rightarrow 4\frac{\delta H_2[u]}{\delta x}$.}. In
the presence of this term, implementation of the {\it deformed} `equation of motion' (EOM) ({\it i. e.}, the KdV equation),

\bea
&&u_t+u_{xxx}-\frac{2}{3}\partial_x\left(\frac{\delta H_1[u]}{\delta u(x)}+u_{xx}\right)+2uu_x=0,\nonumber\\
{\rm or},\quad&&u_t+6uu_x+u_{xxx}={\cal X}_x,\label{14}
\eea
leaves the curvature non-zero. 

\paragraph*{}One can construct an infinite number of quasi-conserved charges through the Abelianization procedure applied
in Ref.s \cite{FZ,FZ1,FZ2}, through gauge-transforming the Lax components. In doing so, the anomaly ${\cal X}$ prevents
rotation of {\it both} of them into the same infinite dimensional Abelian subalgebra of the characteristic $sl(2)$ loop
algebra, eventually leading to an infinite set of quasi-conservation laws characterized by ${\cal X}$. In doing so, to
project-out the anomaly term beforehand, one can perform an initial gauge transformation, which is generically defined
as,

\be
(A,B)\rightarrow U(A,B)U^{-1}+\partial_{(x,t)}UU^{-1}, \quad F_{tx}\rightarrow UF_{tx}U^{-1},\label{15}
\ee
for a {\it constant} gauge choice,

\be
U=\exp\left(-\frac{1}{2}\sigma_3\right),\label{16}
\ee
so that the second term of the first equation in Eq.s \ref{15} does not contribute presently. On further imposing the
deformed equation of motion, this leads to the {\it on-shell} curvature:

\be
\bar{F}_{tx}=\frac{1}{e}\left[u_t+u_{xxx}-\frac{2}{3}\partial_x\left(\frac{\delta H_1[u]}{\delta u(x)}+u_{xx}\right)+2uu_x\right]\sigma_++{\cal X}\sigma_3\equiv{\cal X}\sigma_3,\label{17}
\ee
where $e$ being the exponential function with unit argument. The KdV equation now can be applied as a particular choice
of gauge has been made. The corresponding `rotated' Lax pair is,

\bea
&&\bar{A}=\frac{1}{e}u\sigma_+-e\sigma_- \quad{\rm and}\nonumber\\
&&\bar{B}=-u_x\sigma_3-\frac{1}{e}\left[u_{xx}-\frac{2}{3}\left(\frac{\delta H_1[u]}{\delta u(x)}+u_{xx}\right)\right]\sigma_++2eu\sigma_-.\label{18}
\eea 

\paragraph*{\it The $sl(2)$ loop algebra:}The $SU(2)$ algebraic structure for the KdV system \cite{1} enables the construction
of an $sl(2)$ loop algebra:

\be
\left[b^n,F_{1,2}^m\right]=2F_{2,1}^{m+n}, \quad \left[F_1^n,F_2^m\right]=\lambda b^{m+n},\label{19}
\ee
consistent with the definitions,

\be
b^n=\lambda^n\sigma_3, \quad F_1^n=\frac{1}{\sqrt{2}}\lambda^n\left(\lambda\sigma_+-\sigma_-\right) \quad{\rm and}\quad F_2^n=\frac{1}{\sqrt{2}}\lambda^n\left(\lambda
\sigma_++\sigma_-\right),\label{20}
\ee
with $\lambda$ being the spectral parameter. Such a structure is essentially same as that in Ref. \cite{FZ2} for
quasi-integrable (QI) NLS systems. This serves as a strong connection between QI deformations of the two
systems, which we will address soon. 

\paragraph*{\it The Gauge Transformation:}On following the general approach utilized in Ref.s \cite{FZ,FZ1,FZ2}, we undertake
another gauge transformation with respect to the operator,

\be
g=\exp\left(\sum_{n=0}^{\infty}G_n\right), \quad{\rm where},\quad G_n=a_1^nF_1^n+a_2^nF_2^n.\label{21}
\ee 
Here the coefficients $a_{1,2}^n$ are to be chosen such that the transformed spatial component of the Lax pair
$\tilde{A}=g\bar{A}g^{-1}+g_xg^{-1}$ depends only on $b^n$s:

\be
\tilde{A}\equiv\sum_n\beta^A_nb^n,\label{22}
\ee 
making it diagonal in the $SU(2)$ basis. On considering the BCH formula:

$$e^XYe^{-X}=Y+[X,Y]+\frac{1}{2!}[X,[X,Y]]+\frac{1}{3!}[X,[X,[X,Y]]]+\cdots,$$
the gauge-transformed spatial component takes the form,

\be
\tilde{A}=G_{n~x}+\bar{A}+[G_n,\bar{A}]+\frac{1}{2!}[G_m,[G_n,\bar{A}]]+\frac{1}{3!}[G_l,[G_m,[G_n,\bar{A}]]]+\cdots,\label{N1}
\ee
where summation is understood over all integers, which are semi-positive. The first few of the individual commutators are,

\bea
[G_n,\bar{A}]&=&\frac{1}{\sqrt{2}e}u\left(a_1^n-a_2^n\right)b^n-\frac{e}{\sqrt{2}}\left(a_1^n+a_2^n\right)b^{n+1},\nonumber\\
\frac{1}{2!}[G_m,[G_n,\bar{A}]]&=&-\frac{1}{\sqrt{2}e}u\left(a_1^n-a_2^n\right)\left(a_1^mF_2^{m+n}+a_2^mF_1^{m+n}\right)+\frac{e}{\sqrt{2}}\left(a_1^n+a_2^n\right)\nonumber\\
&\times&\left(a_1^mF_2^{m+n+1}+a_2^mF_1^{m+n+1}\right),\nonumber\\
\frac{1}{3!}[G_l,[G_m,[G_n,\bar{A}]]]&=&\frac{1}{3\sqrt{2}}\left(a_1^la_1^m-a_2^la_2^m\right)\left[\frac{u}{e}\left(a_1^n-a_2^n\right)b^{l+m+n+1}-e\left(a_1^n+a_2^n\right)b^{l+m+n+2}\right],\nonumber\\
\frac{1}{4!}[G_p,[G_l,[G_m,[G_n,\bar{A}]]]]&=&\frac{1}{6\sqrt{2}}\frac{u}{e}\left(a_1^n-a_2^n\right)\left(a_1^la_1^m-a_2^la_2^m\right)\left(a_1^pF_2^{p+l+m+n+1}+a_2^pF_1^{p+l+m+n+1}\right)\nonumber\\
&-&\frac{e}{6\sqrt{2}}\left(a_1^n+a_2^n\right)\left(a_1^la_1^m-a_2^la_2^m\right)\left(a_1^pF_2^{p+l+m+n+2}+a_2^pF_1^{p+l+m+n+2}\right),\nonumber\\
\vdots&&\label{N2}
\eea
The condition of vanishing the coefficients of $F_{1,2}^n$ leads to the order-by-order relations,

\bea
&&{\cal O}\left(F_1^0\right):\quad a_{1~x}^0=\frac{1}{\sqrt{2}}\frac{u}{e}\left(a_1^0-a_2^0\right)a_2^0-\frac{1}{\sqrt{2}}\left(\frac{u}{e\lambda}+e\right),\nonumber\\
&&{\cal O}\left(F_2^0\right):\quad a_{2~x}^0=\frac{1}{\sqrt{2}}\frac{u}{e}\left(a_1^0-a_2^0\right)a_1^0-\frac{1}{\sqrt{2}}\left(\frac{u}{e\lambda}-e\right),\nonumber\\
&&{\cal O}\left(F_1^1\right):\quad a_{1~x}^1=\frac{1}{\sqrt{2}}\frac{u}{e}\left[\left(a_1^0-a_2^0\right)a_2^1+\left(a_1^1-a_2^1\right)a_2^0\right]-\frac{e}{\sqrt{2}}\left(a_1^0+a_2^0\right)a_2^0,\nonumber\\
&&{\cal O}\left(F_2^1\right):\quad a_{2~x}^1=\frac{1}{\sqrt{2}}\frac{u}{e}\left[\left(a_1^0-a_2^0\right)a_1^1+\left(a_1^1-a_2^1\right)a_1^0\right]-\frac{e}{\sqrt{2}}\left(a_1^0+a_2^0\right)a_1^0,\nonumber\\
&&{\cal O}\left(F_1^2\right):\quad a_{1~x}^2=\frac{1}{\sqrt{2}}\frac{u}{e}\left[\left(a_1^0-a_2^0\right)a_2^2+\left(a_1^1-a_2^1\right)a_2^1+\left(a_1^2-a_2^2\right)a_2^0\right]\nonumber\\
&&\qquad\qquad\qquad\qquad-\frac{e}{\sqrt{2}}\left[\left(a_1^0+a_2^0\right)a_2^1+\left(a_1^1+a_2^1\right)a_2^0\right]\nonumber\\
&&\qquad\qquad\qquad\qquad-\frac{1}{6\sqrt{2}}\frac{u}{e}\Big[2\left(a_1^0a_1^1-a_2^0a_2^1\right)\left(a_1^0-a_2^0\right)a_2^0\nonumber\\
&&\qquad\qquad\qquad\qquad+\left(a_1^0a_1^0-a_2^0a_2^0\right)\left\{\left(a_1^1-a_2^1\right)a_2^0+\left(a_1^0-a_2^0\right)a_2^1\right\}\Big]\nonumber\\
&&\qquad\qquad\qquad\qquad+\frac{e}{6\sqrt{2}}\left(a_1^0a_1^0-a_2^0a_2^0\right)\left(a_1^0+a_2^0\right)a_2^0,\nonumber\\
&&{\cal O}\left(F_2^2\right):\quad a_{2~x}^2=\frac{1}{\sqrt{2}}\frac{u}{e}\left[\left(a_1^0-a_2^0\right)a_1^2+\left(a_1^1-a_2^1\right)a_1^1+\left(a_1^2-a_2^2\right)a_1^0\right]\nonumber\\
&&\qquad\qquad\qquad\qquad-\frac{e}{\sqrt{2}}\left[\left(a_1^0+a_2^0\right)a_1^1+\left(a_1^1+a_2^1\right)a_1^0\right]\nonumber\\
&&\qquad\qquad\qquad\qquad-\frac{1}{6\sqrt{2}}\frac{u}{e}\Big[2\left(a_1^0a_1^1-a_2^0a_2^1\right)\left(a_1^0-a_2^0\right)a_1^0\nonumber\\
&&\qquad\qquad\qquad\qquad+\left(a_1^0a_1^0-a_2^0a_2^0\right)\left\{\left(a_1^1-a_2^1\right)a_1^0+\left(a_1^0-a_2^0\right)a_1^1\right\}\Big]\nonumber\\
&&\qquad\qquad\qquad\qquad+\frac{e}{6\sqrt{2}}\left(a_1^0a_1^0-a_2^0a_2^0\right)\left(a_1^0+a_2^0\right)a_1^0;\nonumber\\
&&\vdots\label{N3}
\eea 
satisfying which, we get the desired surviving coefficients of $b^n$ in Eq. \ref{22} as,

\bea
&&\beta^A_0=\frac{1}{\sqrt{2}}\frac{u}{e}\left(a_1^0-a_2^0\right),\nonumber\\
&&\beta^A_1=\frac{1}{\sqrt{2}}\frac{u}{e}\left(a_1^1-a_2^1\right)-\frac{e}{\sqrt{2}}\left(a_1^0+a_2^0\right)-\frac{1}{3\sqrt{2}}\frac{u}{e}\left(a_1^0a_1^0-a_2^0a_2^0\right)\left(a_1^0-a_2^0\right),\nonumber\\
&&\beta^A_2=\frac{1}{\sqrt{2}}\frac{u}{e}\left(a_1^2-a_2^2\right)-\frac{e}{\sqrt{2}}\left(a_1^1+a_2^1\right)-\frac{1}{3\sqrt{2}}\frac{u}{e}\Big[2\left(a_1^0a_1^1-a_2^0a_2^1\right)\left(a_1^0-a_2^0\right)+\left(a_1^0a_1^0-a_2^0a_2^0\right)\nonumber\\
&&\qquad\quad\times\left(a_1^1-a_2^1\right)\Big]+\frac{e}{3\sqrt{2}}\left(a_1^0a_1^0-a_2^0a_2^0\right)\left(a_1^0+a_2^0\right)+\cdots,\nonumber\\
&&\quad\vdots\label{N4}
\eea
It is clear that all the coefficients $a_{1,2}^n$ can completely be determined by solving Eq.s \ref{N3}, thereby leading 
to complete evaluation of the coefficients $\beta^A_n$ in Eq.s \ref{N4}. Therefore, the rotated Lax component $\tilde{A}$
is completely known. 

\paragraph*{}The same gauge transformation transforms the temporal Lax component $\bar{B}$ to
$\tilde{B}=g\bar{B}g^{-1}+g_xg^{-1}$, with corresponding commutators being,

\bea
[G_n,\bar{B}]&=&2u_xG_n-\frac{1}{\sqrt{2}}\frac{f(u)}{e}\left(a_1^n-a_2^n\right)b^n+2eu\left(a_1^n+a_2^n\right)b^{n+1},\nonumber\\
\frac{1}{2!}[G_m,[G_n,\bar{B}]]&=&\frac{1}{\sqrt{2}}\frac{f(u)}{e}\left(a_1^n-a_2^n\right)\left(a_1^mF_2^{m+n}+a_2^mF_1^{m+n}\right)-\sqrt{2}eu\left(a_1^n+a_2^n\right)\nonumber\\
&\times&\left(a_1^mF_2^{m+n+1}+a_2^mF_1^{m+n+1}\right),\nonumber\\
\frac{1}{3!}[G_l,[G_m,[G_n,\bar{B}]]]&=&\frac{1}{3}\left(a_1^la_1^m-a_2^la_2^m\right)\Big[\frac{1}{\sqrt{2}}\frac{f(u)}{e}\left(a_1^n-a_2^n\right)b^{l+m+n+1}\nonumber\\
&&\qquad\qquad\qquad\qquad-\sqrt{2}eu\left(a_1^n+a_2^n\right)b^{l+m+n+2}\Big],\nonumber\\
\frac{1}{4!}[G_p,[G_l,[G_m,[G_n,\bar{B}]]]]&=&-\frac{1}{6\sqrt{2}}\frac{f(u)}{e}\left(a_1^n-a_2^n\right)\left(a_1^la_1^m-a_2^la_2^m\right)\nonumber\\
&&\times\left(a_1^pF_2^{p+l+m+n+1}+a_2^pF_1^{p+l+m+n+1}\right)+\frac{\sqrt{2}}{6}eu\left(a_1^n+a_2^n\right)\nonumber\\
&&\times\left(a_1^la_1^m-a_2^la_2^m\right)\left(a_1^pF_2^{p+l+m+n+2}+a_2^pF_1^{p+l+m+n+2}\right);\nonumber\\
&&\vdots\label{N5}\\
{\rm where},\quad f(u)&=&u_{xx}-\frac{2}{3}\left(\frac{\delta H_1[u]}{\delta u(x)}+u_{xx}\right)\equiv u_{xx}-{\cal X}+2u^2.\nonumber
\eea
The general form of the final temporal component is,

\be
\tilde{B}=\sum_n\left[\beta_n^Bb^n+\varphi^1_nF_1^n+\varphi^2_nF_2^n\right],\label{N6}
\ee
wherein, a few of the coefficients are,

\bea
&&\beta_0^B=-u_x-\frac{1}{\sqrt{2}}\frac{f(u)}{e}\left(a_1^0-a_2^0\right),\nonumber\\
&&\beta_1^B=-\frac{1}{\sqrt{2}}\frac{f(u)}{e}\left(a_1^1-a_2^1\right)+\sqrt{2}eu\left(a_1^0+a_2^0\right),\nonumber\\
&&\beta_2^B=-\frac{1}{\sqrt{2}}\frac{f(u)}{e}\left(a_1^2-a_2^2\right)+\sqrt{2}eu\left(a_1^1+a_2^1\right)+\frac{1}{3\sqrt{2}}\frac{f(u)}{e}\Big[2\left(a_1^0a_1^1-a_2^0a_2^1\right)\left(a_1^0-a_2^0\right)\nonumber\\
&&\qquad\quad+\left(a_1^0a_1^0-a_2^0a_2^0\right)\left(a_1^1-a_2^1\right)\Big]-\frac{\sqrt{2}}{3}eu\left(a_1^0a_1^0-a_2^0a_2^0\right)\left(a_1^0+a_2^0\right)+\cdots,\nonumber\\
&&\quad\vdots\label{N7}\\
&&\varphi_0^1=a_{1~t}^0-\left(\frac{1}{\sqrt{2}}\frac{f(u)}{e\lambda}+\sqrt{2}eu\right)+2u_xa_1^0,\nonumber\\
&&\varphi_1^1=a_{1~t}^1+2u_xa_1^1+\frac{1}{\sqrt{2}}\frac{f(u)}{e}\left[\left(a_1^0-a_2^0\right)a_2^1+\left(a_1^1-a_2^1\right)a_2^0\right]-\sqrt{2}eu\left(a_1^0+a_2^0\right)a_2^0,\nonumber\\
&&\varphi_2^1=a_{1~t}^2+2u_xa_1^2+\frac{1}{\sqrt{2}}\frac{f(u)}{e}\left[\left(a_1^0-a_2^0\right)a_2^2+\left(a_1^1-a_2^1\right)a_2^1+\left(a_1^2-a_2^2\right)a_2^0\right]\nonumber\\
&&\qquad\quad-\sqrt{2}eu\left[\left(a_1^1+a_2^1\right)a_2^0+\left(a_1^0+a_2^0\right)a_2^1\right]+\frac{1}{3\sqrt{2}}eu\left(a_1^0a_1^0-a_2^0a_2^0\right)\left(a_1^0+a_2^0\right)a_2^0\nonumber\\
&&\qquad\quad+\frac{1}{6\sqrt{2}}\frac{f(u)}{e}\Big[2\left(a_1^0a_1^1-a_2^0a_2^1\right)\left(a_2^0-a_1^0\right)a_2^0+\left(a_1^0a_1^0-a_2^0a_2^0\right)\nonumber\\
&&\qquad\qquad\qquad\qquad\quad\Big\{\left(a_2^1-a_1^1\right)a_2^0+\left(a_2^0-a_1^0\right)a_2^1\Big\}\Big]+\cdots,\nonumber\\
&&\quad\vdots\label{N8}\\
&&\varphi_0^2=a_{2~t}^0-\left(\frac{1}{\sqrt{2}}\frac{f(u)}{e\lambda}-\sqrt{2}eu\right)+2u_xa_2^0,\nonumber\\
&&\varphi_1^2=a_{2~t}^1+2u_xa_2^1+\frac{1}{\sqrt{2}}\frac{f(u)}{e}\left[\left(a_1^0-a_2^0\right)a_1^1+\left(a_1^1-a_2^1\right)a_1^0\right]-\sqrt{2}eu\left(a_1^0+a_2^0\right)a_1^0,\nonumber\\
&&\varphi_2^2=a_{2~t}^2+2u_xa_2^2+\frac{1}{\sqrt{2}}\frac{f(u)}{e}\left[\left(a_1^0-a_2^0\right)a_1^2+\left(a_1^1-a_2^1\right)a_1^1+\left(a_1^2-a_2^2\right)a_1^0\right]\nonumber\\
&&\qquad\quad-\sqrt{2}eu\left[\left(a_1^1+a_2^1\right)a_1^0+\left(a_1^0+a_2^0\right)a_1^1\right]+\frac{1}{3\sqrt{2}}eu\left(a_1^0a_1^0-a_2^0a_2^0\right)\left(a_1^0+a_2^0\right)a_1^0\nonumber\\
&&\qquad\quad+\frac{1}{6\sqrt{2}}\frac{f(u)}{e}\Big[2\left(a_1^0a_1^1-a_2^0a_2^1\right)\left(a_2^0-a_1^0\right)a_1^0+\left(a_1^0a_1^0-a_2^0a_2^0\right)\nonumber\\
&&\qquad\qquad\qquad\qquad\quad\times\Big\{\left(a_2^1-a_1^1\right)a_1^0+\left(a_2^0-a_1^0\right)a_1^1\Big\}\Big]+\cdots,\nonumber\\
&&\quad\vdots\label{N9}
\eea
 \paragraph*{}Finally, upon the second transformation, the curvature takes the form,

\bea
&&\tilde{F}_{tx}:=g\bar{F}_{tx}g^{-1}=\tilde{A}_t-\tilde{B}_x+\left[\tilde{A},\tilde{B}\right]\nonumber\\
&&\qquad\equiv{\cal X}\Big[b^0-2\left(a_1^nF_2^n+a_2^nF_1^n\right)-\left(a_1^na_1^m-a_2^na_2^m\right)b^{m+n+1}\nonumber\\
&&\qquad\qquad+\frac{2}{3}\left(a_1^na_1^m-a_2^na_2^m\right)\left(a_1^lF_2^{l+m+n+1}+a_2^lF_1^{l+m+n+1}\right)\nonumber\\
&&\qquad\qquad+\frac{1}{6}\left(a_1^na_1^m-a_2^na_2^m\right)\left(a_1^pa_1^l-a_2^pa_2^l\right)b^{p+l+m+n+2}\nonumber\\
&&\qquad\qquad-\frac{1}{15}\left(a_1^na_1^m-a_2^na_2^m\right)\left(a_1^pa_1^l-a_2^pa_2^l\right)\left(a_1^qF_2^{q+p+l+m+n+2}+a_2^qF_1^{q+p+l+m+n+2}\right)\nonumber\\
&&\qquad\qquad-\frac{1}{90}\left(a_1^na_1^m-a_2^na_2^m\right)\left(a_1^pa_1^l-a_2^pa_2^l\right)\left(a_1^ra_1^q-a_2^ra_2^q\right)b^{r+q+p+l+m+n+3}\nonumber\\
&&\qquad\qquad+\cdots\Big]\nonumber\\
&&\qquad:={\cal X}\sum_n\left(f_0^nb^n+f_1^nF_1^n+f_2^nF_2^n\right),\label{N17}
\eea
with coefficients,

\bea
&&f_0^0=1,\nonumber\\
&&f_0^1=-\left(a_1^0a_1^0-a_2^0a_2^0\right),\nonumber\\
&&f_0^2=-2\left(a_1^0a_1^1-a_2^0a_2^1\right)+\frac{1}{6}\left(a_1^0a_1^0-a_2^0a_2^0\right)^2,\nonumber\\
&&f_0^2=-2\left(a_1^0a_1^2-a_2^0a_2^2\right)-\left(a_1^1a_1^1-a_2^1a_2^1\right)\nonumber\\
&&\qquad+\frac{2}{3}\left(a_1^0a_1^0-a_2^0a_2^0\right)\Big\{\left(a_1^0a_1^1-a_2^0a_2^1\right)-\frac{1}{60}\left(a_1^0a_1^0-a_2^0a_2^0\right)^2\Big\},\nonumber\\
&&\quad\vdots\label{N18}\\
&&f_{1,2}^0=-2a_{2,1}^0,\nonumber\\
&&f_{1,2}^1=2\left[-a_{2,1}^1+\frac{1}{3}\left(a_1^0a_1^0-a_2^0a_2^0\right)a_{2,1}^0\right],\nonumber\\
&&f_{1,2}^1=2\Big[-a_{2,1}^2+\frac{2}{3}\left(a_1^0a_1^1-a_2^0a_2^1\right)a_{2,1}^0+\frac{1}{3}\left(a_1^0a_1^0-a_2^0a_2^0\right)a_{2,1}^1\nonumber\\
&&\qquad\qquad-\frac{1}{30}\left(a_1^0a_1^0-a_2^0a_2^0\right)^2a_{2,1}^0\Big],\nonumber\\
&&\quad\vdots\label{N19}
\eea

\subsection{Quasi-conservation}\label{2.3}In order to demonstrate the deviation from integrability, based on the QI deformation,
it is pertinent to evaluate quantities which would have represent conservation or have themselves be conserved for the
undeformed system. For brevity, contributions of zeroth order in the spectral parameter $\lambda$, which has effectively 
been the order of expansion following the loop algebra defined in Eq. \ref{20}, are considered. With this motive, we
first single-out the available relations from the expressions of different coefficients of that order, obtained in above.
On subtracting the zeroth order terms in Eq.s \ref{N3}, one finds,

\be
a_{-~x}^0+\frac{1}{\sqrt{2}e}u\left(a_-^0\right)^2+\sqrt{2}e=0,\quad{\rm where}\quad a_-^0:=a_1^0-a_2^0,\label{N14}
\ee
which can exactly be solved for a given solution $u(x)$. Further, on considering expressions of $\tilde{A}$, $\tilde{B}$
and $\tilde{F}_{tx}$ from Eq.s \ref{22}, \ref{N6} and \ref{N17} in the zeroth order in the spectral parameter
$\lambda$, the definition of the curvature leads to,

\bea
&&\frac{1}{\sqrt{2}e}\left(ua^0_-\right)_t+u_{xx}+\frac{1}{\sqrt{2}e}\left(f(u)a_-^0\right)_x={\cal X} \quad{\rm and}\nonumber\\
&&\varphi_{0~x}^-+\sqrt{2}\frac{u}{e}\varphi_0^-a_-^0=2{\cal X}a_-^0;\quad{\rm where}\quad\varphi_0^-:=\varphi_0^1-\varphi_0^2,\label{N20}
\eea
wherein the expressions from Eq.s \ref{N4}, \ref{N7}, \ref{N8}, \ref{N9}, \ref{N18} and \ref{N19} have respectively been
utilized. Finally, from Eq.s \ref{N8} and \ref{N9} one finds,

\be
a_{-~t}^0=\varphi_0^-+2\sqrt{2}eu-2u_xa_-^0.\label{N21}
\ee
From Eq.s \ref{N14}, \ref{N20} and \ref{N21}, the system can completely be solved up to the zeroth order. 

\paragraph*{}One can construct $n$ number of quasi-continuity expression for the present system, by considering the
expressions in Eq.s \ref{22} and \ref{N6} followed by the final one in Eq. \ref{N17}. Then, the co-efficients of $b^n$
in the definition of curvature leads to the  of the quasi-continuity forms,

\be
\Gamma^n:=\beta^A_{n~t}-\beta^B_{n~x}\equiv {\cal X}f_0^n,\label{N10}
\ee
which vanish for the undeformed KdV system as ${\cal X}=0$. From the zeroth order ($n=0$) expressions in Eq.s \ref{N4}
and \ref{N7}, 

\be
\Gamma^0=\beta^A_{0~t}-\beta^B_{0~x}=\frac{1}{\sqrt{2}e}\left(ua_-^0\right)_t+u_{xx}+\frac{1}{\sqrt{2}e}\left(f(u)a_-^0\right)_x\equiv{\cal X},\label{N11}
\ee
with the final equality coming from the first of Eq. \ref{N20}. At higher orders, it can be shown that the RHS is a pure
function of the anomaly ${\cal X}$ that vanishes for the undeformed system. Therefore, the anomaly function obtained is
solely responsible for the deviation of the system from integrability, as was observed in Ref. \cite{FerrSG,FZ,FZ1,FZ2}.
It will be shown later, by considering a particular form of deformation, that this deviation is infinitesimal in
magnitude. For the mean-time, from an order-by-order verification, it is easy to foresee that,

\be
\Gamma^n\equiv{\cal X}f_0^n,\label{N26}
\ee
with coefficients $f_0^n$ are given in Eq. \ref{N18}.

\paragraph*{}Following the treatment for QI NLS system in Ref. \cite{FZ2}, on the basis of the weak equivalence of NLS 
and KdV systems \cite{5}, the zeroth order quasi-conserved charge is, 

\be
Q^0:=\int_x\beta_0^A\equiv\frac{1}{\sqrt{2}e}\int_x ua_-^0,\label{N16}
\ee
which needs to be conserved for quasi-integrability.This is consistent with the definition of the quasi-continuity
expression in Eq. \ref{N11} \cite{FZ2}. From the first of Eq.s \ref{N20}, 

\be
\frac{dQ^0}{dt}=\frac{1}{\sqrt{2}e}\int_x \left(ua_-^0\right)_t=\int_x\left[{\cal X}-u_{xx}-\frac{1}{\sqrt{2}e}\left(f(u)a_-^0\right)_x\right]\equiv\int_x{\cal X},\label{N27}
\ee
modulo vanishing total derivatives of functions of the amplitude $u$, that and its derivatives can safely be assumed to
vanish asymptotically. In general,

\be
\frac{dQ^n}{dt}=\int_x{\cal X}f_0^n\equiv\int_x\Gamma^n:=\Lambda^n.\label{N28}
\ee
The RHS above is not zero in general, but the integral can vanish, following judicious choice of ${\cal X}$. Such a result
will physically imply that the asymptotically the system is integrable, which will correspond to the scattering states of
the same. For a localized solution $u$, it is safe to assume that it attains at least a constant value (mostly 0) in such
scattering states, and its derivatives duly vanish. This amounts for the 'asymptotic integrability' of the QI systems
\cite{FZ,FZ1,FZ2}. In the
present case, where the QI deformation is obtained through that of the Hamiltonian $H_1[u]$, a suitable choice for the 
same can yield the desired result. The integrand itself can also exactly be zero, for particular value of $n$, provided
corresponding $f_0^n$ vanishes. This can very much be possible for particular KdV amplitude $u$, leading to a particular
value of $f_0^n$ for which the space-integral of $\Gamma^n$ vanishes.

\section{Comparison with QI NLS System and Definite Results}\label{Sec3}
In order to understand the possibility of explicit quasi-integrability of KdV system, we now try to 
utilize the $\mathbb{Z}_2$ symmetry of $sl(2)$ loop algebra, as explicated in Ref. \cite{FZ2}. We start with equivallently
re-defining the generators as,

\bea
&&\left[b^n,F_{1,2}^m\right]=2F_{2,1}^{m+n}, \quad \left[F_1^n,F_2^m\right]=\delta b^{m+n},\quad{\rm where},\nonumber\\
&&b^n=\lambda^n\sigma_3, \quad F_1^n=\frac{1}{\sqrt{2}}\lambda^n\left(\delta\sigma_+-\sigma_-\right) \quad{\rm and}\quad F_2^n=\frac{1}{\sqrt{2}}\lambda^n\left(\delta
\sigma_++\sigma_-\right),\label{N31}
\eea
with $\delta$ is a parameter, to be identified later. In case of NLS system, it was the sign of the coefficient of the 
non-linear term in the EOM \cite{FZ2}. Under this definition, instead of the constant gauge choice of Eq. \ref{16},
we choose a new one,

\be
\mathfrak U=\exp\left[\tanh^{-1}\left(\frac{\delta+u}{\delta-u}\right)\right]b^0.\label{N32}
\ee
This leads to a new spatial Lax component,

\be
\bar{\cal A}=\left[\tanh^{-1}\left(\frac{\delta+u}{\delta-u}\right)\right]_xb^0-2i\sqrt{\frac{u}{2\delta}}F_2^0.\label{N32}
\ee

The crucial difference between the components $\bar{A}$ of Eq. \ref{18} or $\tilde{A}$ of Eq. \ref{22} and $\bar{\cal A}$
is that only the last one is a definite eigenfunction of the $\mathbb{Z}_2$ transformation mentioned above. It is a combination
of the order 2 automorphism of $sl(2)$ loop algebra:

\be
\Sigma(b^n)=-b^n,\quad\Sigma(F_1^n)=-F_1^n\quad{\rm and}\quad\Sigma(F_2^n)=F_2^n,\label{N33}
\ee
and parity:

\be
{\cal P}: \quad (\tilde{x},\tilde{t})\to(-\tilde{x},-\tilde{t});\quad{\rm with}\quad \tilde{x}=x-x_0 \quad{\rm and}\quad \tilde{t}=t-t_0
\ee
about a particular point $(x_0,t_0)$ in space-time, which can very well be chosen to the origin. These transformations are
mutually commute, as they work in two different spaces ({\it i. e.}, group and coordinate subspaces). Thus,

\be
\Omega\left(\bar{\cal A}\right)=\bar{\cal A},\quad\Omega=\Sigma{\cal P},\label{N34}
\ee
for $u$ being parity-even. It is sensible enough to assume so as KdV equation is parity-invariant to begin with, and the
quasi-modified one in Eq. \ref{14} is also the same, especially subjected to the explicit form of deformation to be 
introduced in the next section\footnote{Practically it amounts to having ${\cal X}_x$ odd in derivatives, which it is.}.
More intuitively, as QI systems support single-soliton structures of the undeformed systems, the well-known bright and 
dark soliton solutions of standard KdV are parity-even. Therefore, it is sensible to consider $u$ as such.

\paragraph*{}The use of $sl(2)$ symmetry was motivated \cite{FZ2} by determining parity properties of the integrands
$\Gamma^n$, as the vanishing of its integral over space (Eq. \ref{N28}) ensures conservation of the corresponding charge
$Q^n$, rendering the deformed system QI. For this purpose, the kernel and image subspaces of $sl(2)$ were considered, with
generators $b^n$ being the semi-simple element that splits the loop algebra into them. To practically utilize this, another
gauge transformation can be performed with respect to,

\be
\mathfrak g=\exp\sum_{n=1}^\infty G^{-n},\quad G^{-n}=\eta_1^{-n}F_1^{-n}+\eta_2^{-n}F_2^{-n},\label{N35}
\ee
which leads to,

\be
\bar{\cal A}\to\tilde{\cal A}={\mathfrak g}\bar{\cal A}{\mathfrak g}^{-1}+{\mathfrak g}_x{\mathfrak g}^{-1},\label{N36}
\ee
with the goal of obtaining a $\mathfrak g$ with definite transformation under $\Omega$. For this purpose, following Ref.
\cite{FZ2}, terms with different powers of spectral parameter $\lambda$ of $\tilde{\cal A}=\sum_m\tilde{\cal A}^{(m)}$
are compared with those in the RHS of Eq. \ref{N36} as,

\bea
&&\tilde{\cal A}^{(0)}=\bar{\cal A},\nonumber\\
&&\tilde{\cal A}^{(-1)}=\left[G^{-1},\bar{\cal A}\right]+\partial_xG^{-1},\nonumber\\
&&\tilde{\cal A}^{(-2)}=\left[G^{-2},\bar{\cal A}\right]+\frac{1}{2!}\left[G^{-1},\left[G^{-1},\bar{\cal A}\right]\right]+\frac{1}{2!}\left[G^{-1},\partial_xG^{-1}\right]+\partial_xG^{-2},\nonumber\\
&&\qquad\quad\vdots\label{N37}
\eea 
However, unlike that for QI NLS system \cite{FZ2}, the spatial Lax component is entirely ${\cal O}(\lambda^0)$, barring
separation of commutators with $b^0$ from those with $F_2^0$ of $\bar{\cal A}$ on the basis of powers of the spectral
parameter. As $b^0$ splits the subspaces, the above fact does not allow to separately obtain $\Omega$-transformations of
the group elements $G^{-n}$ of $\mathfrak g$. This essentially obstructs the exact determination of parity properties of
$\Gamma^n$, unlike that in Ref. \cite{FZ2}. More directly, and decisively, the spatial KdV Lax operator itself
cannot even be confined to the Kernel subspace, by demanding it to be an $\Omega$-eigenstate beforehand. It is confirmed
by the first of the Eq.s \ref{N37}, that requires vanishing of the coefficient of $b^0$ in $\bar{\cal A}$ itself, which
requires the inconsistency $u=0$!, as can be seen from Eq. \ref{N32}.

\paragraph*{}Although surely the KdV and very expectedly the QI KdV are parity-invariant equations, the very nature of
the Lax formulation of the same (Eq. \ref{6}) prohibits a sub-algebraic separation, as they are strictly ${\cal O}(\lambda^0)$.
A direct approach, as in the previous section, can lead to explicit quasi-integrability, through brute-force evaluation
of the coefficients $f_0^n$s and/or through judicious choice of the Hamiltonian $H_1[u]$, as discussed following Eq.
\ref{N28}. The anomaly function ${\cal X}$ for QI NLS \cite{FZ2} is explicitly parity-odd, resulting into vanishing of 
its integral, yielding a conserved charge. For the present case, however, the very definition of the anomaly through
deformation of the Hamiltonian provides a different opportunity. If that deformation yields an anomaly ${\cal X}$ which
is a total derivative of a function of $u$ and its derivatives, by virtue of Eq. \ref{N27}, the charge $Q^0$ will be
conserved embodying qasi-integrability. For an extended (deformed) Hamiltonian,

\be
H_1[u]\to H_1[u]=\int_{-\infty}^\infty\left[\frac{1}{2}u_x^2-u^3+\epsilon F(u)\right],\label{N38}
\ee
wherein $\epsilon$ is the deformation parameter, we consider the example $F(u)=\frac{3}{4}uu_{xx}$. This will immediately
yield ${\cal X}=\epsilon u_{xx}$ (Eq. \ref{13}) yielding a conserved $Q^0$, {\it irrespective} of the parity or any other
properties of $u$. From Eq. \ref{14}, or from the definition of EOM with the Hamiltonian,

$$u_t=\left[\frac{\delta H[u]}{\delta u}\right]_x,$$
the QI KdV equation takes the form,

\be
u_t+6uu_x+(1-\epsilon)u_{xxx}=0,\label{N39}
\ee
which is essentially a scaling of the undeformed system, supporting similar solutions, including solitons. This is expected %%%%%%%%%%%%%%%%%%%%%%%%%%%%%%%%%%%%%%%%%
as the choice for $F(u)$ is nothing but a total derivative away from the first term in $H_1[u]$, modulo $\epsilon$.
Therefore it will yield eventually a completely integrable system, with single soliton solutions of the form,

\be
u=\frac{c}{2}{\rm sech}^2\left[\sqrt{\frac{c}{4(1-\epsilon)}}\left(x-ct-x_0+ct_0\right)\right],\quad c>0,\label{Sol1}
\ee
with speed $c$. Non-trivially and more importantly, however, this provides an opportunity to construct a hierarchy of 
{\it higher-derivative} extensions of KdV, with different choices of $F(u)$. For demonstration, we consider the
following two:

\be
F(u)=-\frac{3}{2m}\epsilon u_x^m \quad{\rm and}\quad F(u)=\frac{3}{4}\epsilon uu^{(2n)}, \quad{\rm with}\quad m=3,4,\cdots;\quad n=1,2,\cdots,\label{N40}
\ee
where $m$ is ordinary power and $n$ is the order of space derivatives, leading to the higher-derivative EOMs,

\bea
&&u_t+6uu_x+u_{xxx}=\epsilon\left(u_x\right)_{xx}^{m-1}\quad{\rm and}\nonumber\\
&&u_t+6uu_x+u_{xxx}=\epsilon u^{(2n+1)},\label{N41}
\eea
respectively. A numerous other possibilities are there and we aspire to analyze such QI systems in the future. It would
be interesting study their solitonic structures, as solutions like that in Eq. \ref{Sol1} do not satisfy them.

\paragraph*{}Thus far, although KdV is parity-symmetric, the particular algebraic structure of the system does not allow
for a direct utility of that property to be incorporated into the corresponding QI analysis\footnote{
One may directly construct a $\mathfrak g$ with parity-odd $\eta_1^{(-n)}$ and parity-even $\eta_2^{(-n)}$, or
vice versa. However, such an approach is analytically not practical. However, in principle, it confirms that {\it parity-odd}
$f_0^n$s can be constructed, as in Ref. \cite{FZ2}, which shall lead to quasi-conservation, as ${\cal X}$ is usually
parity-even for a parity-even $u$, as seen in the discussion preceding immediately. But the very fact that a parity-definite
spatial Lax component has not been consistently obtained, makes this possibility very bleak.}. However, the proposed deformation
of the Hamiltonian can be utilized to obtain a hierarchy of more extensive QI systems, with ${\cal X}$ itself being a total space-derivative,
ensuring at least one conserved charge ($Q^0$). The fact that at present it is not sure whether these systems support
solitonic structures, or even definite solutions, motivates further studies. In the next section, we will consider a
perturbation approximation of the QI deformation for KdV, utilizing a known weak-coupling mapping to NLS system.

\section{Infinitesimal QI Deformation: Weak-coupling NLS Analogy}\label{3}
Now we consider {\it infinitesimal} deformation of the KdV Hamiltonian $H_1[u]$ in the spirit of Ref.s \cite{FZ,FZ1,FZ2}.
Therein, the non-linear term in $u$ can be deformed as $u\rightarrow u^{1+\epsilon},\quad 0<\epsilon\ll 1$, to yield,

\be
H^{\rm Def}_1[u]\equiv\int_{-\infty}^\infty dx~\left(\frac{1}{2}u_x^2-u^{3+3\epsilon}\right).\label{N22}
\ee
This leads to the expression of the anomaly term,

\be
{\cal X}\approx-2\epsilon u^2\left(1+3\log(u)\right),\label{N23}
\ee
which is ${\cal O}(\epsilon)$ as expected, as $u$ is finite in general. It is clear that the above anomaly does not posses
a definite parity and will not conform to a conserved charge ($Q^0$ at least). However, in this particular case, the
expectation of quasi-integrability arise from the asymptotic behavior of the QI KdV amplitude $u$ \cite{FZ,FZ1}, and
thereby that of the anomaly. This is because, following Eq. \ref{14}, it is the {\it derivative} of ${\cal X}$ that 
appears in the EOM and a localized $u$, especially under weak coupling, corresponds to $u_x\to 0$ for $\vert x\vert\to\infty$. 
As claimed in subsection \ref{2.3}, this clearly makes RHS of Eq. \ref{N28} to vanish in the same limit, approaching
integrability, as expected.

\subsection{Relation with NLS system}
The choice of such a deformation arise from the similar amplitude deformation to attain QI Non-linear
Schr\"odinger equation \cite{FZ2}. These two systems can be mapped at the {\it solution level} under the  {\it weak-coupling}
approximation, expressed as \cite{5},

\bea
&&u_{\rm NLS}=\varepsilon\left(\varphi e^{i\theta}+\bar{\varphi}e^{-i\theta}\right)+\frac{\varepsilon^2}{k_0^2}\left(\varphi^2 e^{i2\theta}+\bar{\varphi}^2e^{-i2\theta}\right)-2\frac{\varepsilon^2}{k_0^2}\vert\varphi\vert^2; \quad{\rm where},\label{31}\\
&&\theta=k_0x+\omega_0t, \quad 0<\varepsilon\ll 1, \quad \omega_0=k_0^3\neq 0.\nonumber
\eea
Substituting the above mapping in the KdV equation \ref{8}, and comparing terms of ${\cal O}\left(\varepsilon^3\right)$
and with phase $e^{i\theta}$, one arrives at the NLS equation,

\be
\varphi_T+i3k_0\varphi_{XX}+i\frac{6}{k_0}\vert\varphi\vert^2\varphi=0,\label{32}
\ee
and its complex conjugate,

\be
\bar{\varphi}_T-i3k_0\bar{\varphi}_{XX}-i\frac{6}{k_0}\vert\varphi\vert^2\bar{\varphi}=0,\label{33}
\ee
for phase $e^{-i\theta}$, with respect to the new coordinates,

\be
T=\varepsilon^2t \quad{\rm and}\quad X=\varepsilon\left(x+3k_0^2t\right).\label{34}
\ee
In Eq.s \ref{32} and \ref{33}, the `time'-derivative term comes from that of the KdV, the second derivative term comes
from the third derivative term of the same, and the non-linear term comes from its counterpart in KdV. Such direct correspondence,
though approximate, motivates enough for adopting the present approach to obtain QI KdV system. However, one should expect
that the present deformation works as long as the mapping of Eq. \ref{31} persists, implying {\it another} infinitesimal
parameter $\varepsilon$.

\paragraph*{}From Eq. \ref{14}, the proposed QI deformation of the KdV amplitude directly leads to the infinitesimally
deformed KdV equation,

\be
u_t+u_{xxx}+\left(6+10\epsilon+12\epsilon\log(u)\right)uu_x=0,\label{35}
\ee
by considering terms up to ${\cal O}(\epsilon)$. As $u$ is ${\cal O}(\varepsilon)$, the third term in the above bracket
is sub-dominant than the second, and hence, can be dropped out. Then it can straight-forwardly be shown that the proposed
KdV deformation invariably leads to QI NLS system. From Eq.s \ref{31} and \ref{35}, the modified NLS equation is obtained
as\footnote{In the map of Eq. \ref{31} \cite{5}, the non-linear terms of KdV and NLS systems map exclusively to each-other.
Therefore, any scaling of the one in the KdV equation, like that in Eq. \ref{35} when the logarithm is neglected,
corresponds to the same scaling of the similar term in the NLS equation.},

\be
\varphi_T+i3k_0\varphi_{XX}+i\left(1+\frac{5}{3}\epsilon\right)\frac{6}{k_0}\vert\varphi\vert^2\varphi=0.\label{36}
\ee 
It is easy to see that,

\bea
&&\left(1+\frac{5}{3}\epsilon\right)\vert\varphi\vert^2\approx\frac{\delta}{\delta\vert\varphi\vert^2}V(\vert\varphi\vert) \quad{\rm with},\nonumber\\
&&V(\vert\varphi\vert)\equiv\frac{1}{2}\vert\varphi\vert^{4\left(1+\tilde{\epsilon}\right)}, \quad{\rm where}, \tilde{\epsilon}=\frac{5}{3}\epsilon,\label{37}
\eea
following the physical fact that the density $\vert\varphi\vert^2$ is sufficiently small in the weak-coupling limit. 
Therefore, the particular deformation the KdV Hamiltonian in Eq. \ref{N22}, under the weak-coupling approximate
mapping of Ref. \cite{5}, leads to the QI deformation of NLS system, given by Eq. \ref{37}, which of the
same form as given in Ref. \cite{FZ2}. This intuitively ensures that as long as the mapping prevails, the deformation of
Eq. \ref{N22} leads to QI KdV.

\paragraph*{}Therefore, the justification of constructing a QI KdV system in the lines of the QI NLS system \cite{FZ2}
is quite valid. The construction of the $sl(2)$ loop algebra in Eq. \ref{20} being essentially being the same is thus
justified, with consistent QI charges obtained in Eq. \ref{N28}. 

\paragraph*{Specific Solutions:}The one-soliton solution for the quasi-deformed KdV equation in Eq. \ref{35}, after
dropping-out the logarithm, has the form,

\be
u=\frac{c}{2}{\rm sech}^2\left[\frac{1}{2}\sqrt{c\beta}\left(x-x_0-\beta c(t-t_0)\right)\right],\quad{\rm where}\quad \beta=1+\frac{5}{3}\epsilon,\quad c>0,\label{Sol2}
\ee
which is again the standard one, with parameter/variable scaling, as the approximate deformation amounts for the scaling
of the non-linear term. The corresponding one-soliton solution for the NLS system of Eq. \ref{36} is,

\bea
&&\varphi(X,T)=K{\rm sech}\left[\Lambda_1K\left(\Lambda_2\tilde{X}-V\tilde{T}\right)\right]\exp\left[\frac{i}{2}\Lambda_2V\tilde{X}+\frac{i}{4}\left(\Lambda_1^2K^2-V^2\right)\tilde{T}\right];\label{Sol3}\\
&&{\rm where}\quad\Lambda_1=i\sqrt{\frac{3\beta}{k_0}},\quad\Lambda_2=-\frac{i}{\sqrt{3k_0}},\qquad\tilde{X}=X-X_0,\quad\tilde{T}=T-T_0,\nonumber\\
&&{\rm and}\quad (K,V,X_0,T_0)\in{\mathbb R}_+\otimes{\mathbb R}\otimes{\mathbb R}\otimes{\mathbb R},\nonumber
\eea
which again displays similar parameter/variable scaling. The incorporation of the first non-trivial (local) contribution
in the KdV sector, by bringing the logarithm in Eq. \ref{35} back, is of the form $v^2v_x$, with
$u=1+v,\quad v\in{\mathbb R}_-$. For small and positive $u$, $v\lesssim 1$ is relatively large in magnitude. However, 
such an addition of mKdV-type non-linearity may not lead to any conservation in general. However, by introducing a 
{\it simultaneous} deformation of the type discussed in section \ref{Sec3}, namely $F(u)\equiv F(v)\sim v^4$, such a 
system can be reduced to KdV again, in terms of $v$ now. Equivalently, for a simultaneous $F(u)\equiv F(v)\sim v^3$,
in can become a `pure' mKdV system in $v$, with well-known solitonic profiles. However, it should be re-stressed that the
preceding discussion is valid only under weak-coupling approximation, that validates the KdV-NLS mapping of Ref. \cite{5},
further allowing $u$ to be small (but not tending to zero) owing to the `weak' non-linearity.

\subsection{Connection with Non-Holonomic Deformation}
It is fruitful to compare the results obtained thus far with those of NH deformation of KdV and NLS systems. The NH deformation
is practically obtained through extending the Lax components with local terms negative in power of the spectral parameter
$\lambda$. This induces an inhomogeneous extension to the original PDE, with additional local constraints imposed-on the 
deformation parameters, obtained though retainment of the zero-curvature condition. This leaves the deformed system still
{\it integrable}. The NH deformation had been well-analyzed for KdV and coupled complex KdV systems \cite{1}, from both loop-algebraic  
and AKNS approaches, and it has recently been shown in case of NLS systems \cite{AGM} that the NH deformation is essentially
different from QI deformation, as the latter leaves the system non-integrable. The QI deformation is done usually at the 
level of functions of the variable \cite{FZ,FZ1,FZ2,N1} or functionals, as in the present case for KdV, that deforms the 
Lax component itself without extending it spectrally. This necessarily yields a non-zero curvature, whose vanishing leads
to that of the deformation itself. Therefore, both the deformations are fundamentally different. However, in the approximate
regime, such as in the present section, the QI deformation is perturbative in nature, with asymptotic regaining of integrability.
This is so as the $\log(u)$ term is relatively negligible at that limit, aided by the weak coupling, yielding a KdV system
with constant parametric scaling. This is same as was observed for NLS system in Ref. \cite{AGM}.

\paragraph*{}In the same asymptotic limit, this `weak' QI deformation can be identified with a particular version of NH
deformation, having local coefficients that satisfy order-by-order relations in power of the now-small QI-parameter
$\epsilon$, that are identified with constraints. This is intuitively supported, as the QI systems show asymptotic integrability
supporting single and multiple soliton solutions. It will be interesting to identify such systems with order-by-order 
relations (`constraints') by evaluating asymptotic form of the exact solution for QI KdV and other systems.

\section{Deformation of General Complex Coupled KdV Equations}
On considering the Lax pair for coupled complex KdV system \ref{1}, the QI deformation can be accommodated as,

\be
B=\left(\bar{q}q_x-q\bar{q}_x\right)\sigma_3-\left[\bar{q}_{xx}-\frac{2}{3}q\left(\frac{\delta\bar{H}_1\left[\bar{q}\right]}{\delta\bar{q}}+\bar{q}_{xx}\right)\right]\sigma_++\left[q_{xx}+\frac{2}{3}\bar{q}\left(\frac{\delta H_1[q]}{\delta q}+q_{xx}\right)\right]\sigma_-.\label{26}
\ee
This leads to the curvature,

\bea
&&F_{tx}=\left[\bar{q}_t+\bar{q}_{xxx}-\frac{2}{3}\left(q\frac{\delta\bar{H}_1\left[\bar{q}\right]}{\delta\bar{q}}+q\bar{q}_{xx}\right)_x+2\vert q\vert^2\bar{q}_x-2q_x\bar{q}^2\right]\sigma_+\nonumber\\
&&\qquad\quad-\left[q_t+q_{xxx}+\frac{2}{3}\left(\bar{q}\frac{\delta H_1[q]}{\delta q}+\bar{q}q_{xx}\right)_x-2\vert q\vert^2q_x+2\bar{q}_xq^2\right]\sigma_-\nonumber\\
&&\qquad\quad+{\cal X}_c\sigma_3,\label{27}
\eea
where, the anomaly function now reads as,

\be
{\cal X}_c\equiv\frac{2}{3}\bar{q}^2\left(\frac{\delta H_1[q]}{\delta q}+q_{xx}\right)+\frac{2}{3}q^2\left(\frac{\delta\bar{H}_1\left[\bar{q}\right]}{\delta\bar{q}}+\bar{q}_{xx}\right).\label{28}
\ee
A gauge-transformation, similar to Eq. \ref{16}, leads to the {\it on-shell} curvature upon using the deformed equation
of motion as,

\bea
&&\bar{F}_{tx}=\frac{1}{e}\left[\bar{q}_t+\bar{q}_{xxx}-\frac{2}{3}\left(q\frac{\delta\bar{H}_1\left[\bar{q}\right]}{\delta\bar{q}}+q\bar{q}_{xx}\right)_x+2\vert q\vert^2\bar{q}_x-2q_x\bar{q}^2\right]\sigma_+\nonumber\\
&&\qquad\quad-e\left[q_t+q_{xxx}+\frac{2}{3}\left(\bar{q}\frac{\delta H_1[q]}{\delta q}+\bar{q}q_{xx}\right)_x-2\vert q\vert^2q_x+2\bar{q}_xq^2\right]\sigma_-\nonumber\\
&&\qquad\quad+{\cal X}_c\sigma_3\nonumber\\
&&\qquad\equiv{\cal X}_c\sigma_3,\label{29}
\eea
along-with the corresponding Lax pair:

\bea
&&\bar{A}=\frac{1}{e}\bar{q}\sigma_+-eq\sigma_- \quad{\rm and}\nonumber\\
&&\bar{B}=\left(\bar{q}q_x-q\bar{q}_x\right)\sigma_3-\frac{1}{e}\left[\bar{q}_{xx}-\frac{2}{3}q\left(\frac{\delta\bar{H}_1\left[\bar{q}\right]}{\delta\bar{q}}+\bar{q}_{xx}\right)\right]\sigma_+\nonumber\\
&&\qquad+e\left[q_{xx}+\frac{2}{3}\bar{q}\left(\frac{\delta H_1[q]}{\delta q}+q_{xx}\right)\right]\sigma_-.\label{30}
\eea
On comparing Eq.s \ref{29} and \ref{30} with Eq.s \ref{17} and \ref{18} respectively for the real KdV system, it is easy
to obtain the replacements,

\bea
&&{\cal X}\rightarrow{\cal X}_c;\nonumber\\
&&\frac{u}{e}\rightarrow\frac{\bar{q}}{e} \quad{\rm and}\quad e\rightarrow eq \quad{\rm in}\quad \bar{A};\nonumber\\
&&u_x\rightarrow q\bar{q}_x-\bar{q}q_x,\quad \frac{1}{e}f(u)\rightarrow\frac{1}{e}\bar{F}\left(q,\bar{q}\right) \quad{\rm and}\quad 2eu\rightarrow eF\left(q,\bar{q}\right) \quad{\rm in}\quad \bar{B};\label{N24}
\eea
wherein,

\be
\bar{F}\left(q,\bar{q}\right):=\bar{q}_{xx}-\frac{2}{3}q\left(\frac{\delta\bar{H}_1\left[\bar{q}\right]}{\delta\bar{q}}+\bar{q}_{xx}\right) \quad{\rm and}\quad F\left(q,\bar{q}\right):=q_{xx}+\frac{2}{3}\bar{q}\left(\frac{\delta H_1[q]}{\delta q}+q_{xx}\right).\label{N25}
\ee
for the complex coupled system. These replacements can directly be mapped in the equations of subsection \ref{2.3}, and
the corresponding generalized results can be obtained. Without going into the details of the calculations, we can write down the
extended versions of equations \ref{N14}, \ref{N20} and \ref{N21} respectively as,

\bea
&&a_{-~x}^0+\frac{\bar{q}}{\sqrt{2}e}\left(a_-^0\right)^2+\sqrt{2}eq=0,\nonumber\\
&&\frac{1}{\sqrt{2}e}\left(\bar{q}a_-^0\right)_t+\left(\bar{q}q_{xx}-q\bar{q}_{xx}\right)+\frac{1}{\sqrt{2}e}\left(\bar{F}a_-^0\right)_x={\cal X}_c,\nonumber\\
&&\varphi_{0~x}^-+\frac{\sqrt{2}}{e}\bar{q}a_-^0\varphi_0^-=2{\cal X}_ca_-^0 \quad{\rm and}\nonumber\\
&&a_{-~t}^0=\varphi_0^-+2\sqrt{2}eF+2\left(q\bar{q}_x-\bar{q}q_x\right)a_-^0.\label{N30}
\eea
This enables us directly to obtain the quasi-conservation equations,

\be
\frac{dR^n}{dt}\equiv\int_x{\cal X}_cf_0^n,\label{N29}
\ee 
where $R^n$ is the corresponding anomalous charge of the $n$-th order. In general, the corresponding $\tilde{F}_{tx}$
coefficients will have the same {\it algebraic form} as those in Eq.s \ref{N18} and \ref{N19}, as no direct $u$-dependence ($q$ and $\bar{q}$
dependence in this case) appear therein. However, the explicit expressions will be extended following the replacements of
Eq.s \ref{N25}, though it is always possible to obtain a pair of $\left(q,\bar{q}\right)$ yielding $f_0^n=0$ for a particular
value of $n$. This ensures the quasi-integrability of complex coupled KdV system, as per subsection \ref{2.3} and
section \ref{3}.

\paragraph*{}On a more specific note, following the treatment in Sec. \ref{Sec3}, the expression of the anomaly in Eq.
\ref{28} leaves even a wider choice of deformations of the {\it two} Hamiltonians $\bar{H}_1[\bar{q}]$ and $H_1[q]$,
to construct a ${\cal X}_c$ as a total space-derivative. Then, as the second of Eq.s \ref{N30} ensures that
$\frac{dQ^0}{dt}=\int_x{\cal X}_c$ as before, one can again construct a scaled coupled complex KdV system, as well as
a tower of QI, higher-derivative extended KdV models, which can very well be of a wider interest.

\section{Conclusions}
It is seen that a quasi-integrable deformation of the KdV system is indeed possible, provided the loop-algebraic generalization
\cite{1} has been considered. Further, as the KdV equation does not represent `dynamics', in the sense of neither Galilean
(like NLS) nor Lorentz (like SG) systems, the deformation has to be performed at an {\it off-shell} level ({\it i. e.},
without using the EOM). The available Hamiltonian formulation of KdV system comes to rescue in this respect, wherein both
local extensions of the Hamiltonian density and power-deformation of the corresponding amplitude have been adopted. The
prior allows for constructing scaled KdV, with single-soliton profile, as well as families of higher-derivative extensions
to the same, with at least one conserved charge. The latter is intuitively allowed, following the weak correspondence
between KdV and NLS systems, and the compatibility of the present deformation with that of QI NLS system has been obtained,
followed by corresponding one-soliton solutions with variable scaling. It will be interesting to obtain and analyze
particular stable solutions to this deformed KdV and higher derivative systems, and to study their behavior with those from
QI NLS system when the weak correspondence is valid. We aspire to imply the same for complex coupled KdV formalism.

\vskip 0.5cm
\noindent{\it Acknowledgement:} PG is grateful to Jun Nian and Vasily Pestun for interesting discussions. The authors
are also grateful to Professors Luiz. A. Ferreira, Wojtek J. Zakrzewski and Betti Hartmann for their encouragement,
various useful discussions and critical reading of the draft.

\end{document}